\documentclass[letterpaper,onecolumn]{article}
\usepackage[utf8]{inputenc}
\usepackage{amsmath}
\usepackage{amsfonts}
\usepackage{amssymb}
\usepackage{graphicx}
\begin{document}
\author{Matei Radulescu}
\title{Why does the detonation cell size $\lambda$ provide quasi-universal scaling of detonation dynamics?}

\author{{\large Matei Ioan Radulescu$^{a}$}\\[10pt]
        {\footnotesize \em $^a$Department of Mechanical Engineering, University of Ottawa, Ottawa, Canada}}

\date{}  


\maketitle
\begin{abstract} 
The detonation cell size $\lambda$ is known to provide a characteristic scale for the detonation dynamics and serves as scaling parameter for numerous correlations with other dynamic parameters of detonations, e.g., the critical detonation tube diameter for transmission to open space of $d\simeq13\lambda$ across many reactive systems.  In the present paper, we address the reason why such scaling works so well.  We formulate a macroscopic reaction rate model for cellular detonations relying on the distribution of ignition delays in cellular detonations, conditioned by $\lambda$.  The ignition delay distribution is modeled in terms of the cellular dynamics, using the closed form model recently proposed by Radulescu (JFM 2025) for the lead shock dynamics in a unit cell.  The distribution of ignition delays obtained in closed form serves as a simple model for the phase-averaged reaction rate in a cell-averaged description.  The global reaction rate model is applied to the problem of globally curved cellular detonations.  Excellent agreement with experiment is obtained.  The paper thus highlights for the first time a simple dynamic closure for cellular detonations and provides the simple answer why the cell size provides the relevant length scale for detonations:  it provides the length of the mean detonation structure, or its hydrodynamic thickness. 
\end{abstract}

\section{Introduction\label{sec:introduction}} 
\begin{figure}
\centering
\includegraphics[width=1\textwidth]{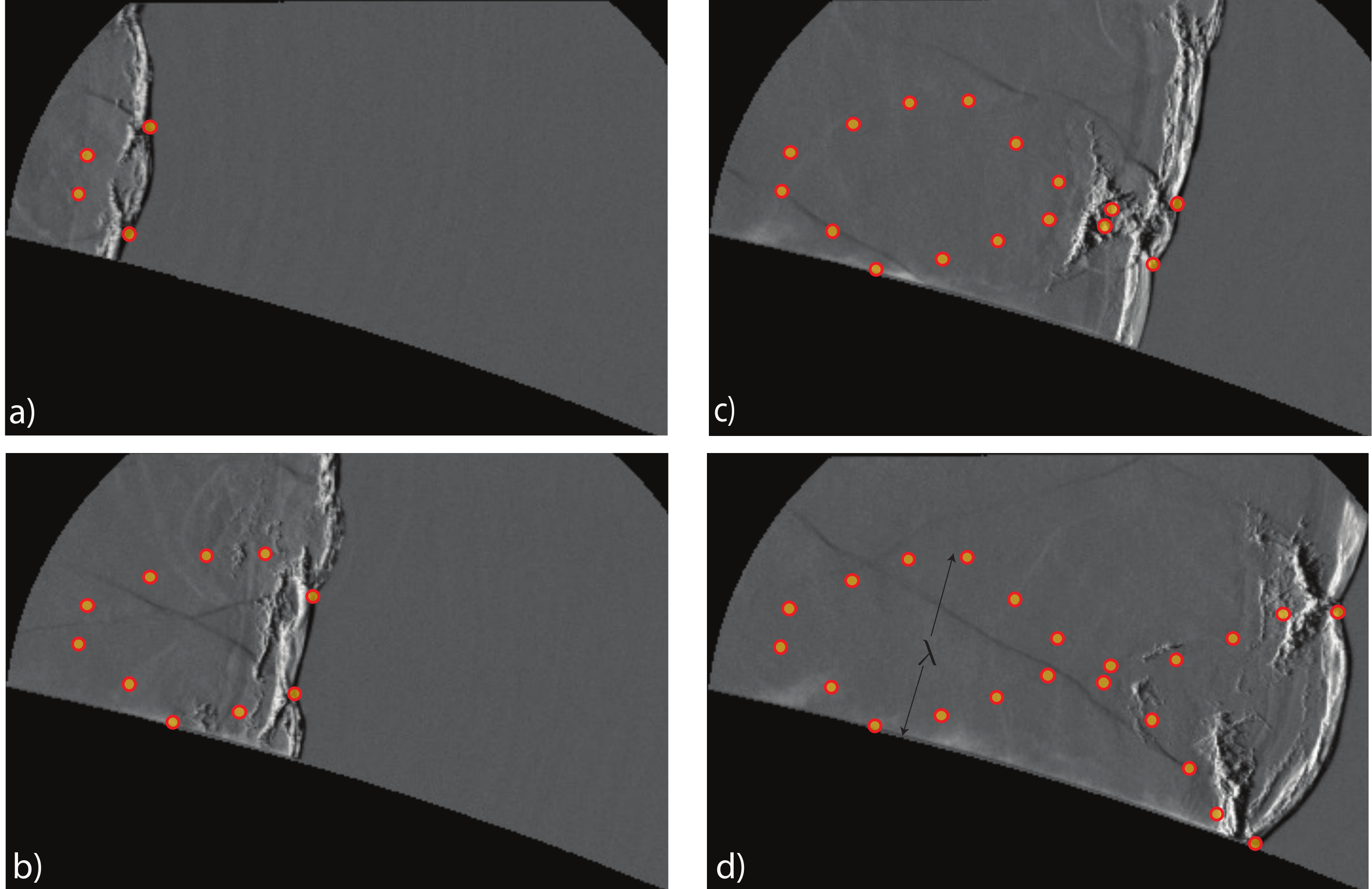}
\caption{Globally curved cellular detonation in CH$_4$ + 2O$_2$ illustrating the accumulation of non-reacted gas behind the decaying lead shock and its subsequent consumption  by fast flames on the surface pockets; adapted from Xiao et al \cite{xiao2020role}.  Orange points denote positions of triple points registered from the sequential frames. Note that between frames \textit{c} and \textit{d}, the non-reacted pocket accumulated from the previous cell is consumed after the transverse wave collisions.}
\label{fig:qiang1}
\end{figure}

Gaseous detonations have a cellular structure.  The front displays a self-generated non-steady structure with cells taking on a preferred transverse dimension $\lambda$.  An example of the cellular dynamics obtained in our laboratory in a relatively thin channel with an exponentially enlarging cross-section is shown in Figure \ref{fig:qiang1}.  The lead fronts are always decaying and gases accumulate as non-reacted behind the front; the ignition typically quenches before the shock gets to the mid-cell location due to the volumetric expansion cooling behind decaying shocks  \cite{lundstrom1969,cheevers2022ignition,frederick2024}.  Gases do however re-ignite from triple shock reflections, which are responsible for increasing the local temperature sufficiently for the gases accumulating as non-reacted gas behind the front to react very rapidly.  In the example of Fig.\ \ref{fig:qiang1}, the non-reacted pockets are consumed by fast flames on their surface \cite{xiao2020role}. Other instances show the immediate reaction of all gas accumulated more rapidly by direct shock ignition behind the transverse waves, as recently documented by Frederick et al. \cite{Frederick_Gejji_Shepherd_Slabaugh_2025}.  Experiments thus suggest that the gases react after transverse wave collisions, and the cellular time provides approximately the longest time for all the gases processed by a unit cell to react \cite{radulescu2025meanP}.   The cell size thus provides a characteristic length scale for the global reaction zone thickness.

Since the discovery of the detonation cellular structure in the late 1950's by Denisov and Troshin \cite{denisov1959}, numerous correlations have been developed using $\lambda$ to form the basis of scaling laws for the detonation dynamics; see for example Lee's survey \cite{lee1984}, which remains the state of the art to this day.  For example, the critical detonation tube diameter for transmission to open space is $d\simeq13\lambda$ for many reactive systems. The basis of these correlations were never clarified and only few efforts have been attempted using artificial blast wave models of the cellular dynamics \cite{vasiliev, hu2002analytical, crane2021}.  The main objective of the present study is to develop a physics informed model linking the detonation dynamics with the cellular structure.

\begin{figure}
\centering
\includegraphics[width=0.7\textwidth]{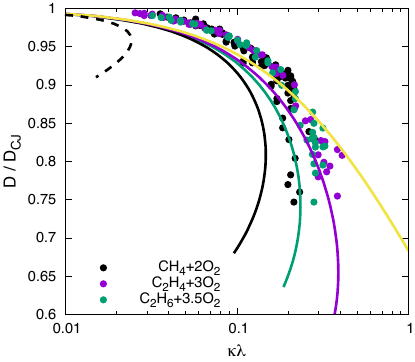}
\caption{Detonation speed dependence on mean curvature $\kappa \lambda$ obtained experimentally by \cite{xiao2020role}.  Yellow line is the model prediction using the closed form model of the present paper and fixed cellular time.  Color coded lines are model predictions with state sensitive cellular times according to \eqref{eq:tcelldiff} for diffusively controlled pocket burn-out; broken line is the model prediction for CH$_4+2$O$_2$ according to \eqref{eq:tcellshock} for auto-ignition controlled pocket burn-out.}
\label{fig:qiang2}
\end{figure}

The universal scaling with $\lambda$ was demonstrated empirically by Nakayama et al. \cite{nakayama2013} for globally curved detonations, a feature believed to control most detonation losses.  In our laboratory, using channels with cross-sectional area $A$ increasing exponentially of the type shown in Figure \ref{fig:qiang1}, we also isolated waves with a global constant curvature $\kappa=\mathrm{d}\ln A/ \mathrm{d}s$ and determined their quasi-steady structure and speed $D$ \cite{radulescu2018dynamic, xiao2020role}.  When scaling the mean detonation curvature with cell size, the data were found to collapse to a single curve for irregular structure detonations, as shown in Fig.\ \ref{fig:qiang2}. In this paper, we explain why $\lambda$ provides scaling of detonations with mean curvature for irregular structure detonations, detonations argued to be sustained by transverse wave interactions \cite{radulescu2002failure}. In doing so, we formulate a novel paradigm whereby the macroscopic, or phase averaged, reaction rate model for cellular detonations is given by the distribution of ignition delays in cellular detonations, conditioned by $\lambda$.

This short paper is organized as follows.  We first explain how the distribution of ignition delays in the limit of thin reaction zones provides directly a model for the global reaction rate.  We then seek to construct the temporal distribution of energy release in a model cell in which all the gas reacts at the end of the cellular cycle. The global reaction rate model relies on the modeling of the residence time of the gas behind the shock prior to ignition. The averaged rate of energy release is conditioned only by the cellular time, which remains unclosed, and shape of distribution of the lead shock speed. The model is used to reconstruct the hydrodynamic structure of curved cellular detonations, in order to compare with experimental data and assess the merit of the approach.
		
\section{Distribution of ignition delays is equivalent to the global reaction rate\label{sec:1}}

One could ask the question: 

\textit{What is the distribution of ignition delays for all the fluid particles crossing the lead shock in a cellular detonation?}  

Choosing a characteristic "average" cell, the question simplifies to what is the distribution of ignition delay of the gases crossing the shock in a single representative cell.  While this distribution is obtained below, we would first like to highlight the somewhat surprising result (at first glance) that the distribution of ignition delays is equivalent to the phase averaged reaction rate or
\textit{global} reaction rate, such as required in a hydrodynamic description in a Favre averaged approach suggested in the past (see \cite{radulescu2007hydrodynamic} and references therein.)

Consider the probability density function of the ignition delay $\mathrm{pdf} (t_i)= w(t_i)$,  then $w(t_i)\delta t_i$ is the amount of gas that has ignition delay between times $t_i$ and $t_i +\delta t_i$ by definition in a stationary process.  If each particle deposits all its energy in a vanishing time compared to $t_i$, then $Q w(t)\delta t_i$ is also the amount of energy that has been liberated between times  $t$ and $t +\delta t$.  In the limit $\delta t \rightarrow 0$, $w(t)$ is thus the \\textit{global} reaction rate explicitly formulated in terms of time.

An equivalent derivation is to consider the cumulative density function $\mathrm{cdf} (t_i)$, an approach presented in Lalchandani's thesis \cite{lalchandani}.  This represents the mass fraction of fluid, say $Y_b$, that has reacted before the time $t_i$ in a given process, on average. 
\begin{align}
Y_b(t)=\int_0^t \mathrm{pdf}(t_i) \mathrm{d}t_i= \mathrm{cdf}(t_i=t)
\end{align}
 The time derivative of $Y_b$ is by definition the global reaction rate, on average. But so is the derivative of the cumulative density function, which is
\begin{align}
\mathrm{pdf} (t_i) =\frac{\mathrm{d}}{\mathrm{d} t_i}\mathrm{cdf} (t_i)
\end{align}
We thus arrive at the same conclusion that the pdf of ignition delay can provide an adequate closure for the reaction rate in a stationary process. 

\begin{figure}
\centering
\includegraphics[width=0.7\textwidth]{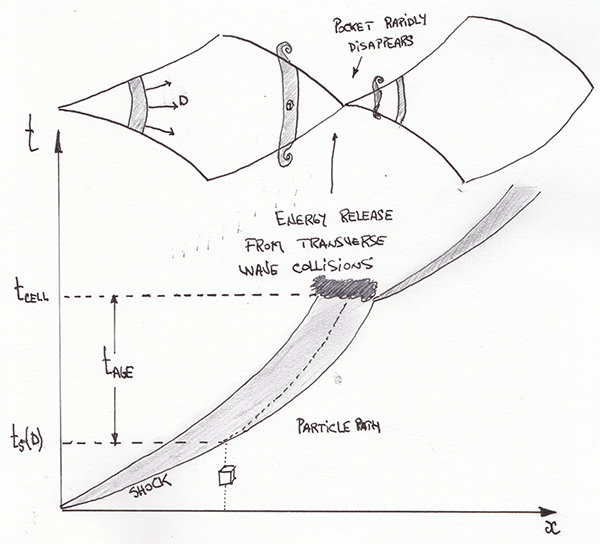}
\caption{The adopted model for rapid ignition of all gases processed by the lead shock at the end of the cellular cycle; the residence time of the gas, or its age $t_{age}$, is the time spent behind the lead shock until the end of the cell when transverse wave collisions induce a rapid reaction, assumed instantaneous in the present model.}
\label{fig:modelsketch}
\end{figure}

\section{$\mathrm{pdf}(D)$ from the cell cycle dynamics}

We model the detonation cell dynamics with the recently proposed model of Radulescu \cite{radulescu2025gasdynamic}, illustrated in Figure \ref{fig:modelsketch}.  In their model, the lead shock is periodically reformed following the energy release of all the gas that has accumulated as non-reacted behind the lead shock since the previous pulse. This model is in accord with phenomenology for irregular detonations, where the gases processed by the lead shock accumulate as non-reacted pockets due to the expansion cooling associated with the rapid shock decay \cite{cheevers2022ignition, Frederick_Gejji_Shepherd_Slabaugh_2025}. 

If one assumes that gas to react very fast at the end of the cell cycle from the temperature increase from transverse interactions, then the dynamics of the lead shock can be obtained in closed form in terms of the heat release in the gas $Q$, ratio of specific heats $\gamma$ and period of the pulsation $\tau$.  In between pulses, the lead shock decays exponentially in time and linearly in space.  See the model of \cite{radulescu2025gasdynamic} for details.

\begin{figure}
\centering
\includegraphics[width=0.7\textwidth]{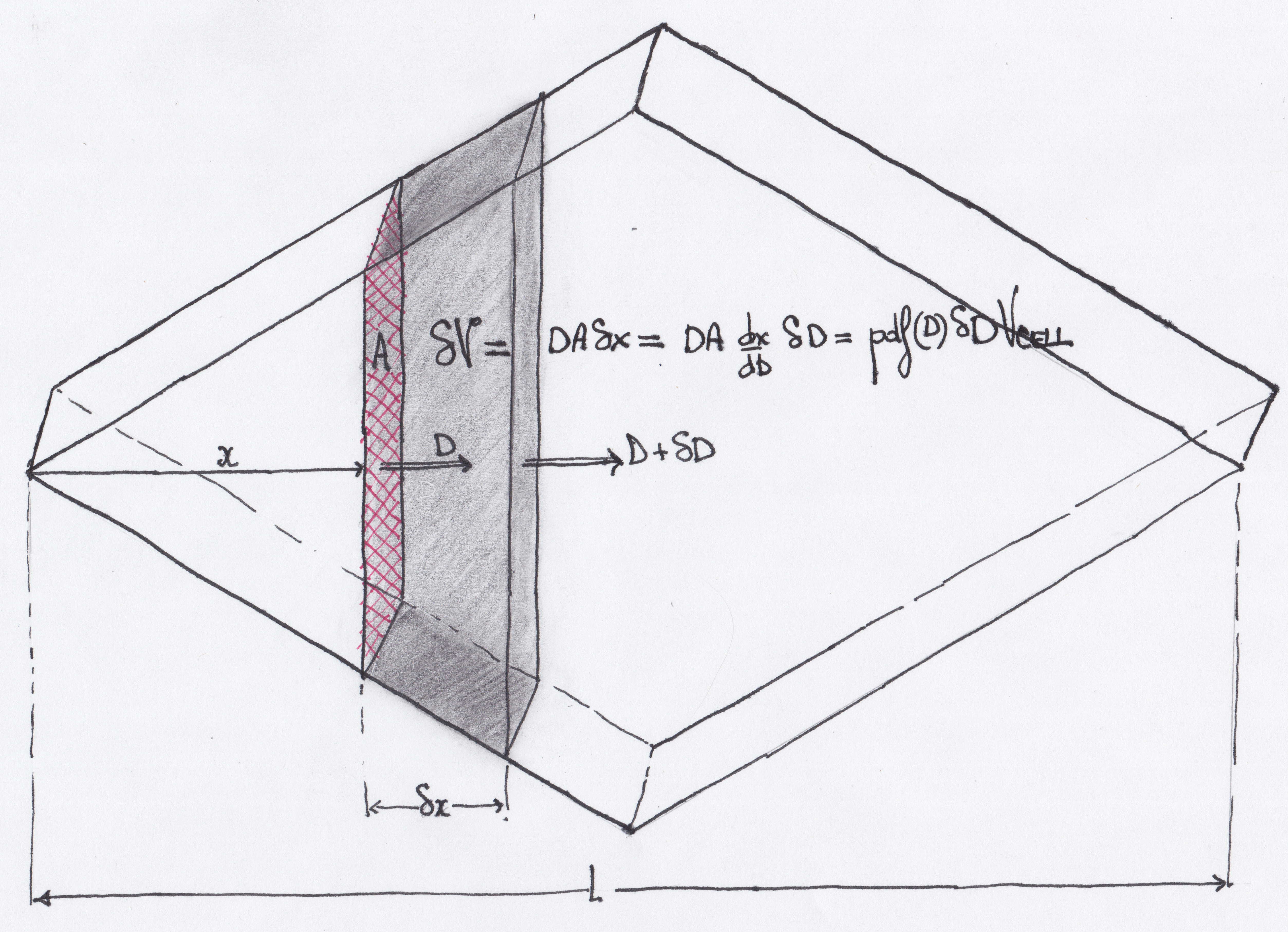}
\caption{Idealized cellular geometry permitting to express the fraction of gas propagating between speeds $D$ and $D + \delta D$ during one average cycle to the volume ($\delta V$) swept by the shock propagating at speed $D$ for a distance  $\delta x = (\mathrm{d} x /\mathrm{d}D) \delta D$ over the cell volume.}
\label{fig:cellconstruct}
\end{figure}

If one assumes that the cells are 2D, the pdf of the lead shock speed can be readily obtained by geometric arguments, given the speed evolution law inside the cell and its geometry, as shown in Figure  \ref{fig:cellconstruct}.  A similar derivation was first communicated in Lalchandani's Masters thesis \cite{lalchandani}. 
A detailed derivation is given in the supplementary material listed as an appendix.
We assume that the speed is continuously decaying, such that there is a monotonic relation $D(x)$ that can be inverted and is single-valued.  

The desired probability $\mathrm{pdf}(D) \delta D$ is, by definition, the fraction of material that gets shocked at speeds ranging between $D$ and $D + \delta D$ during one cycle.  Let $\delta V$ denote the volume swept by the shock propagating at speed $D$ over a distance $\delta x$ in a normal direction, and $\delta V/V_{cell}$ denoting required fraction, where $V_{cell}$ is the total volume swept by the shock in one cell cycle - see Fig.\ \ref{fig:modelsketch}.  We get 
\begin{align}
 \mathrm{pdf}(D) |\delta D| = \frac{\delta V }{V_{cell}}
\end{align}
The swept volume is also $\delta V=A |\delta x|$, where $A$ is the surface area of the shock, so, 
\begin{align}
 \mathrm{pdf}(D) |\delta D| = \frac{A }{V_{cell}}|\delta x|
\end{align}
Since the $\delta D$ and $\delta x$ are related by 
\begin{align}
\delta x = (\mathrm{d} x /\mathrm{d}D) \delta D \label{eq:dddx}
\end{align}
the desired probability density of shock speed is given by
\begin{align}
 \mathrm{pdf}(D) = \frac{A }{V_{cell}}|\frac{\mathrm{d} x }{\mathrm{d}D}| \label{eq:pdf}
\end{align}
We note that the surface area can be parametrized by distance in order to bring out its form $A(D)$ and its influence on the local $\mathrm{pdf}(D)$: large speeds in the early part of the cell are only experienced on small surfaces, while speeds close to the mean are experienced by much larger surfaces. 
 
If the surface area $A(x)$ is linear (the cell is approximated by a parallelogram) and the velocity decay is exponential in time (and linear in space), it can be shown that the pdf of lead shock speed inside a 2D cell is given approximately by the following distribution  
\begin{align}
\mathrm{pdf}(D) \simeq \frac{\Delta D \pm (D-\bar{D} )}{\Delta D^2} \label{eq:pdfD}
\end{align}

\begin{figure}
\centering
\includegraphics[width=0.7\textwidth]{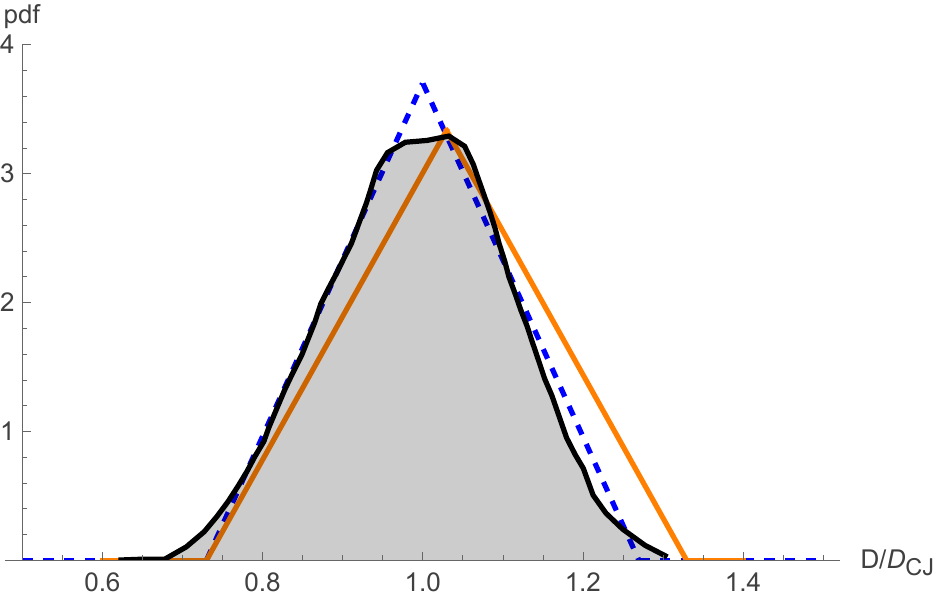}
\caption{Comparison of the analytical shock distribution model \eqref{eq:pdfD} (dashed blue line) for $\beta = \Delta D / \overline{D} = 0.27$ with the experimental data of Frederick et al.\ \cite{frederick2023} in methane-oxygen detonations; orange line is the pdf obtained from \eqref{eq:pdf1} without the assumption of $D_{mid}=\overline{D}$ - see the Appendix for details.}
\label{fig:Fred}
\end{figure}

This result is derived in the supplementary material of the appendix. The upper operator in \eqref{eq:pdfD} (in $\pm$) denotes the pdf for the gases shocked after the mid-cell location and the lower symbol the gases shocked before.  The peak of the pdf occurs at $D=\overline{D}$ and takes the value of $\Delta D^{-1}$. $\Delta D$ is the speed difference between the mean and the end of the cell; we assumed $\Delta D/D <<1$ to arrive at \eqref{eq:pdfD}.  This simple triangular distribution is in very good agreement with Frederick's experiments \cite{frederick2023} , as shown in Fig.\ \ref{fig:Fred}.  
  
If the lead shock speed was not uniform on its surface and the surface was not flat, as assumed in the model, the pdf of the lead shock speed would require a more careful examination requiring the decomposition of the normal lead shock speed into components in a 2D map (or 3D map).  This would be required when reducing experimental data of speeds distributed as a field in a speed map, as introduced by Frederick et al.\ when analyzing experimentally determined speed maps\cite{frederick2023}.
\section{Distribution of non-reacted residence time} 

In our pulsed reaction model, all the gas liberates its energy at the end of the cell cycle only.  When transverse waves collide and the gas reacts, the shocked gas has a non-homogeneous \textit{age} or residence time distribution behind the lead shock.  For example, the gas that has crossed the shock in the early parts of the cell has been shocked for a long time by the time it reacts from transverse wave collisions (approaching the cell time), while the gas shocked at the end of the cell has just been shocked when it reacts from transverse wave collisions.  This is shown schematically in Fig.\ \ref{fig:modelsketch}.  In the limit of our model, the distribution of residence time in a unit cell can be obtained from the shock kinematics alone in closed form, 
as derived in the supplementary material of the appendix. 

\begin{equation}
\mathrm{pdf}(t_{age}) = \frac{2}{\beta t_{cell}}\left(\pm\frac{D}{\bar{D}}\left(t_{age}\right)+\beta\mp1\right)\frac{D}{\bar{D}}\left(t_{age}\right) \label{eq:pdfti}
\end{equation}
with $\beta = \Delta D/\bar{D}$ and 
\begin{equation}
\frac{D}{\bar{D}}=(1-\beta)\exp\left( 2\beta \frac{t_{age}}{t_{cell}}\right)
\end{equation}
The upper operator in \eqref{eq:pdfti} (in $\pm$ and $\mp$) denotes the pdf for the gases shocked after the mid-cell location and the lower symbol the gases shocked before.  
\section{Distribution of residence time as global reaction rate} 
\begin{figure}
\centering
\includegraphics[width=0.7\textwidth]{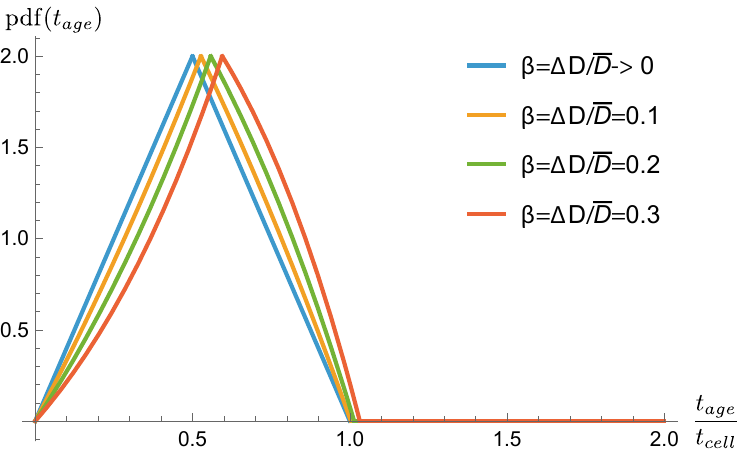}
\caption{The distribution of ignition delay inside a cell given by \eqref{eq:pdfti} for different speed amplitudes $\beta$.}
\label{fig:pdfti}
\end{figure}

The pdf of residence time or \textit{age} serves as a global reaction rate. 
\begin{equation}
w(t')= \mathrm{pdf}(t_{age}=t') \label{eq:pdfti2}
\end{equation}
where $t'$ becomes the time elapsed along a particle path falling behind the mean shock in a quasi-steady description. The expression only takes as parameter the cell time $t_{cell}$ and the speed fluctuation $\beta$. It is the dependence through $t_{cell}$ only that the time scale of energy release affects the dynamics. It is shown in Fig.\ \ref{fig:pdfti}. As can be expected, the global reaction rate for a cellular detonation approaches a triangular shape:  the distribution of residence time behind a steadily moving shock is controlled by the linear variation of the front alone. That the distribution, hence reaction rate is only very weakly affected by the model parameter $\beta$ is the sought after feature that makes the global reaction rate in cellular detonations weakly affected by anything other then the cellular time.   

\begin{figure}
\centering
\includegraphics[width=0.7\textwidth]{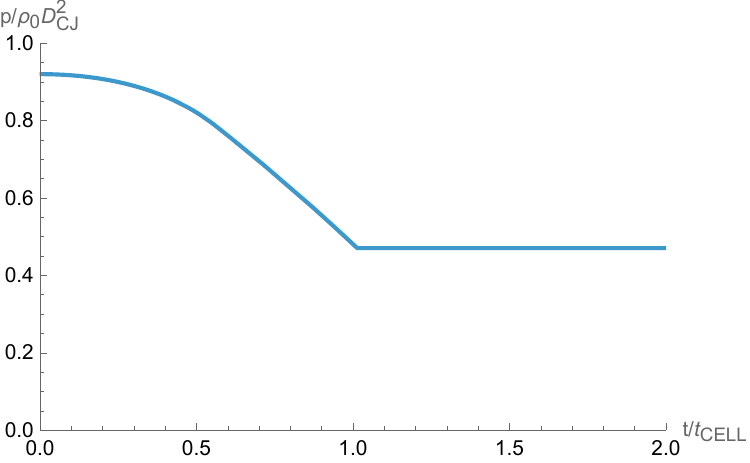}
\caption{The mean pressure evolution in a CH$_4$+2O$_2$ detonation at 6 kPa predicted by the current model.}
\label{fig:znd}
\end{figure}
\section{Global detonation structure}
The global averaged rate of energy addition obtained from the pdf of the ignition  delay can now serve as a macro-kinetic model for detonation waves.  For example, one can ask the question of how the mean pressure decays in a detonation wave.  Using the global reaction rate obtained $w(t) = \mathrm{pdf(t)}$, one can integrate the usual 1D ZND equations (listed below) with the new rate in order to obtain the global  hydrodynamic structure.  Figure \ref{fig:znd} shows the pressure structure for parameters ($\gamma$ = 1.17, $Q/(p_0/\rho_0)$ = 64, $Ea/(p_0/\rho_0)$ = 52 and $\beta = 0.2$) corresponding to a CH$_4$+2O$_2$ detonation at 6 kPa - see below. By construction, the pressure reaches the CJ solution in one mean cell time. 

\begin{figure*}
\centering
\includegraphics[width=1\textwidth]{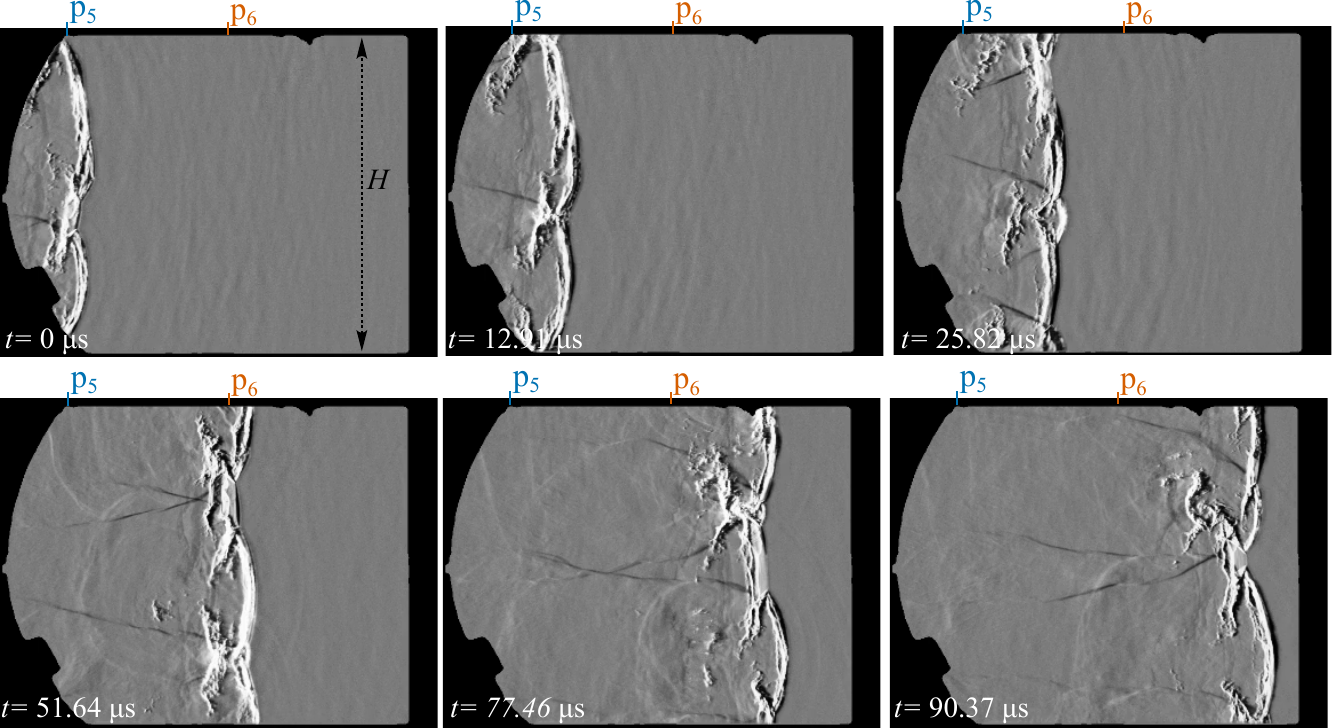}
\caption{The evolution of detonation structure in CH$_4$+2O$_2$ at 6 kPa \cite{radulescu2025meanP}.}
\label{fig:expdet}
\end{figure*}

\begin{figure}
\centering
\includegraphics[width=0.7\textwidth]{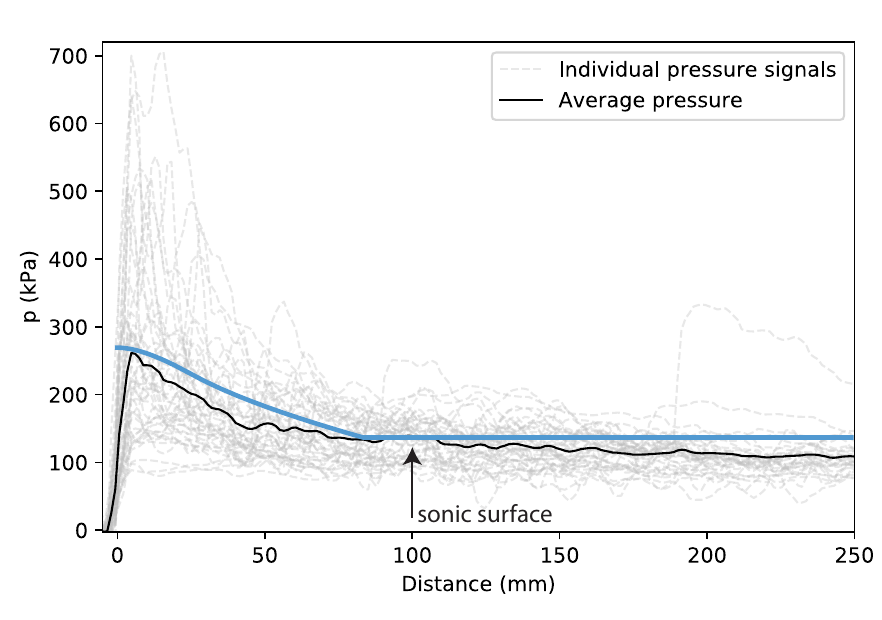}
\caption{The mean pressure evolution in a CH$_4$+2O$_2$ detonation at 6 kPa obtained from ensemble averaging \cite{radulescu2025meanP} and the current model (blue); arrow indicates the position of the sonic surface inferred from the experiments.}
\label{fig:expdet2}
\end{figure}

The predicted pressure evolution can be compared to experiment.  Recently, we established experimentally the mean pressure evolution in detonation cells \cite{radulescu2025meanP}.  Consider for example the detonation wave propagating in CH$_4$+2O$_2$ detonation at 6 kPa in a narrow channel, illustrated in Figure \ref{fig:expdet}.  One can readily identify the largest cell, that takes approximately 80$\mu$s to complete.    The average pressure over approximately 100 traces is given in Fig.\ \ref{fig:expdet2}.  According to the present model, the mean cellular time would be half of the largest time recorded, in order to account for all sizes of cells present, i.e. $t_{cell} \approx 40 \mu$s.  Indeed, for a lack of a better estimate, we assume that the “average” cell size is halfway between zero and the largest one recorded.  This appears a reasonable assumption for irregular detonations, where cells of all sizes are observed all the way to an “outer scale”.  

The model mean pressure evolution is shown superposed on the experimental data in Fig.\ \ref{fig:expdet2}.  The structure predicted by the simple model compares favourably with the experiments.  Note however that shorter ignition delays are expected in the experiments, with a distribution more strongly skewed towards shorter times; indeed, gases are undergoing shorter ignition delays in the early part of the cells, \cite{cheevers2022ignition, Frederick_Gejji_Shepherd_Slabaugh_2025}, the model positing an extreme view that \textbf{all} gases accumulate as non-reacted until the end of the mean cell cycle and ignite from transverse wave collisions. 

\section{Curved detonations}
The true test for the model's value is whether it predicts the global dynamics of detonations with losses and possible extinction.  To demonstrate its usefulness, consider the problem of quasi-steady curved detonations discussed in the introduction.  The steady travelling wave solution is readily obtained by solving the one-dimensional conservation laws with a term appearing to account for the lateral divergence of the flow.  See for example the treatment of Radulescu and Borzou \cite{radulescu2018dynamic}.  The steady solution is obtained when the effect of flow divergence is balanced by the exothermicity at the mean sonic surface \cite{bdzil2012}.  

Briefly, the steady detonation structure is given by the following differential relations, taking the distance behind the lead shock $x$ as independent variable.
\begin{equation}
\frac{dt}{dx}=\frac{1}{u}
\end{equation}
\begin{equation}
\frac{d u}{dx}=\frac{\dot{\sigma}-(\bar{D}-u)\kappa}{\eta} \label{eq:master}
\end{equation}
\begin{equation}
\frac{d p}{dx}=-\rho u \frac{d u}{dx}\
\end{equation}
\begin{equation}
\frac{d \rho}{dx}=-\frac{\rho}{u} \left(\frac{d u}{dx}+(\bar{D}-u)\kappa  \right)
\end{equation}
The global reaction rate $w(t)$ entering in the thermicity expression 
\begin{align}
\dot{\sigma}&=(\gamma-1)\frac{Q}{c^2}w(t)\\
& = (\gamma-1)\frac{Q}{c^2} \mathrm{pdf}(t)
\end{align}
is given by \eqref{eq:pdfti} explicitly in terms of time and model parameters $t_{cell}$ and $\beta$.  
The sonic parameter $\eta$ is 
\begin{equation}
\eta=1 -  \frac{u^2}{c^2}
\end{equation}
The mean curvature of the front is constant in our case, as it is imposed by the global geometrical divergence of the channel 
\begin{equation}
\kappa=\frac{1}{A}\frac{d A}{d s}
\end{equation}
where $s$ is distance in the lab frame of reference in the direction of propagation.  These are the classical ZND equations with flow divergence and their treatment is now well understood \cite{bdzil2012}.  The integration starts at the shock.  With a guessed detonation speed $\bar{D}$, the post shock state provides the conditions at the shock where the integration of the equations begins. If the guessed detonation speed is too low, the subsonic flow with area increase never reaches Mach of unity, as the exothermicity is too slow.  If the guessed speed is too high, the subsonic flow reaches Mach of unity while the flow is still reacting - the rate of reaction is too fast. By virtue of \eqref{eq:master} and division through 0, the gradients tend to infinity.  The sonic solution desired is bracketed by the two scenarios, and corresponds to the solution reaching the sonic solution while the numerator of \eqref{eq:master} vanishes exactly as the sonic point is approached, exothermicity balancing exactly the area change in driving flow, the so-called generalized Chapman-Jouguet solution.  The sonic point being a saddle, the desired solution can be obtained by numerically bisecting the two families of integral curves described with the desired resolution.  Our integration follows this strategy, and is implemented in Mathematica.

The dependence of detonation speed on curvature is obtained in this fashion. Using the rate given by \eqref{eq:pdfti}, the detonation speed is uniquely determined by the parameter $t_{cell}$.  If $t_{cell}$ is assumed constant, i.e., the cellular time is unaffected by velocity deficits, the resulting response is that of Fig.\ \ref{fig:qiang2} (yellow line) and depends very weakly on model gas parameters $\gamma$ and $Q$ .  The agreement with experiment is very good for speeds larger than 0.9$D_{\mathrm{CJ}}$ for all three reactive mixtures.  This clearly highlights that detonations respond to the mean reaction rate effected by the cellular structure:  the cell size providing the characteristic thickness. The agreement with experiment deteriorates as the points of extinction are reached.  

\section{State sensitivity and extinction} 
In the model presented above, the relation between cell time and detonation speed was  absent.  This is of course an idealization, since experiment suggests that cell size is dependent on mean speed.  In the context of the model presented, it is expected that the cell time be proportional to the time required to burn out all the gases processed by the cell.  In irregular detonations, since the pockets burn mostly by diffusive processes (flames) on their surfaces, with shear layers enhanced by Kelvin-Helmholtz instability and further disruption by shock surface interactions leading to Rayleigh-Taylor-like hydrodynamic instabilities coupled with gas expansion from reactions \cite{radulescu2007hydrodynamic}, we thus expect $t_{cell}\propto \lambda/S_T \propto \lambda/S_L$, where $S_T$ is the turbulent burning velocity and $S_L$ is the laminar burning velocity.  Since the laminar burning velocity is primarily controlled by the reaction rate near the burned temperature, i.e., 
\begin{align}
S_L\propto \sqrt{\exp\left(-\frac{E_a}{RT_b} \right)}
\end{align}
and the burned gas temperature is a function of detonation speed via the Rankine-Hugoniot relations ($T_b\propto D^2$ in the strong shock limit), one arrives at the cell time dependence on shock speed:
\begin{equation}
\frac{t_{cell}}{t_{\mathrm{cell,CJ}}}=\exp \left(\frac{E_a}{2RT_{\mathrm{CJ}}} \left(\left(\frac{\bar{D}}{D_{\mathrm{CJ}}} \right)^{-2}-1 \right) \right) \label{eq:tcelldiff}
\end{equation}
Repeating the calculations of the last section with the supplementary state dependence of the cell time provides an excellent account for the experiments, as shown in Fig.\ \ref{fig:qiang2}.  In all instances, the model now predicts extinction and captures the modification of the turning point in each fuel.  Methane having a larger activation energy is more sensitive to temperature changes; detonations extinguish at smaller scaled curvatures then the other fuels.  Note however that the effect is quite weak, the reason stemming from the relevant temperature in the sensitivity analysis being the burned gas temperature for flames, and not the unburned one, for auto-ignition.   

The model formulated can also be used to test other modeling assumptions.  For example, the commonly accepted hypothesis is that the cell time is proportional to the induction time evaluated behind a shock propagating at the CJ speed.   Instead of \eqref{eq:tcelldiff}, one arrives at 
\begin{equation}
\frac{t_{\mathrm{cell}}}{t_{\mathrm{cell,CJ}}}=\exp \left(\frac{E_a}{RT_{\mathrm{VN}}} \left(\left(\frac{\bar{D}}{D_{\mathrm{CJ}}} \right)^{-2}-1 \right) \right)\label{eq:tcellshock}
\end{equation}
This provides a much higher sensitivity of the cell time to speed owing to the absence of the square root dependence for flames and the lower characteristic temperature, and yields a prediction of extinction incompatible with experiment, as shown in Fig.\ \ref{fig:qiang2}.  Indeed, as noted by Radulescu, cellular detonations with irregular structures do not notably change the cell size even when velocity deficits are very substantial \cite{radulescu2007hydrodynamic}. Future work should thus address how the cell size is affected by losses and speed deficits - as it provides a clear clue with respect to burning mechanism.

In spite of the model's extreme simplicity, we are able to capture the experimentally established dynamics of diverging detonations very well.  It is evident that the model under-predicts the true rate of energy release in cells, as auto-ignition events are responsible for gas ignition in the early parts of the cells, approximately 1/3 along the cell cycle \cite{cheevers2022ignition,Frederick_Gejji_Shepherd_Slabaugh_2025}.  

Accounting for the early ignition in the cells would skew the triangular distribution illustrated in Figure 6 towards shorter times. A simple model would be the assumption of a quasi-steady ignition in the early part of the cell and application of the current model for the gases escaping auto-ignition behind the lead shock, in the remaining part.  Closure to obtaining the cell size and distribution of energy release of course requires establishing the location of the cut-off and the role of transverse wave compression.  An interesting suggestion made by Vidal is that cut-off is expected at the mid cell location \cite{MONNIER2023112996}, suggested by symmetry arguments.  This has empirical support for more regular systems \cite{lundstrom1969, JACKSON20132260, cheevers2022ignition}.  Under the experimentally supported constraint that all gas reacts within a cell time and shock dynamics of the type assumed here, such modifications to the current model can, in principle, provide closure to both determining the cell size in a given reactive mixture and inferring the global rate of energy release conditioned by the cell.  This effort is under-way and will be communicated in the near future.  

\begin{figure}
\centering
\includegraphics[width=0.7\textwidth]{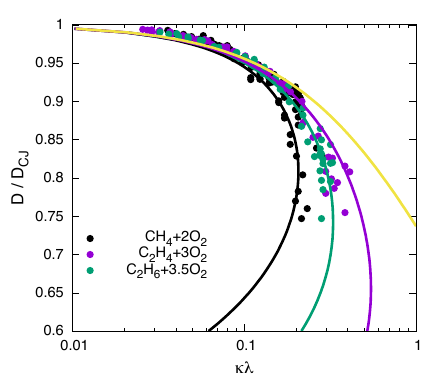}
\caption{Detonation speed dependence on mean curvature $\kappa \lambda$ obtained experimentally by \cite{xiao2020role}.  Model lines same as Fig.\ \ref{fig:qiang2} with the rate being increased by a factor of 1.4.}
\label{fig:corrected}
\end{figure}

Empirically, we also find that artificially enhancing the rate predicted by the model by a factor of 1.4 modifies the model predictions and yields excellent agreement with experiment (Fig.\ \ref{fig:corrected}). This gives us hope that a refinement of the ignition delay distribution model along the lines presented can improve predictions.  However, these are not warranted before the cell size itself becomes better defined in terms of its distribution and more accurate methods to measure it are developed. As advocated by Lee, perhaps the hydrodynamic thickness itself is a better measure \cite{lee1984}; our mean pressure evolution experiments being one method to establish it. 

In more regular systems where the cell size and inner dynamics are more easily measured, the dynamics of curved detonations are also well collapsed by scaling the global curvature by the cell size, as indicated in Figure 18 of \cite{xiao2020role}.  The difference being a factor of 2 with the more "irregular" detonations.  This factor of two is absorbed by the factor of two required in the present model for defining the average cell size in more irregular mixtures.  This brings a non-adjusted model by a factor of 2 in very good agreement with experiment.  Whether this agreement is fortuitous is unclear at present, but is encouraging.  Even for regular systems, auto-ignition is apparently quenched at mid-cell location \cite{lundstrom1969, JACKSON20132260, cheevers2022ignition} and the remaining gas reacts fast by auto-ignition behind reactive transverse waves.  In near limit conditions, the burning mechanism of unreacted pockets escaping ignition by transverse waves also relies on surface flame phenomena, as recently demonstrated by Watanabe et al. \cite{watanabe2026}.  A refinement of the present model for the distribution of ignition delay in more regular systems is clearly worthwhile and efforts are underway.

A long standing debate in the detonation literature - perhaps a tacit one - is whether the cellular detonations should be viewed as a global or local phenomenon.  The present results suggest the mean description, phase averaged over a detonation cell, can provide a meaningful approach. 
 
\section{Outlook and closing remarks}
The present study introduced a framework for modeling cellular detonation waves.  We show that the phase averaged kinetics can be obtained from a model of the ignition delay distribution in a cell. Detonation dynamics governed by such global kinetics were found in excellent agreement with experiment.  This clarifies and provides justification why cell-size scaling works so well in practice.  We also show that the model formulated can also be used to test different hypotheses for the physics responsible for the cellular cycle.  

While future work should provide a more physically accurate model for the ignition distribution in cellular detonations, it would be of interest to further assess if our simple model can be used for predictive purposes in general problems of detonation dynamics. One immediate inquiry is to test the sensitivity of the model to the detonation cell geometry.  In our model, we assumed a purely 2D cell.  In 3D, cells can adopt different configurations \cite{takai}, which will modify the distribution of speeds inside the cell.  It is expected that 3D cells to favour speeds near the mean and have narrower distributions of ignition delays - this can serve to explain the long standing question why 3D cells provide a more robust propagation than 2D cells.

\section*{Acknowledgements}
This work was supported by AFOSR grant FA9550-23-1-0214 (Dr.\ Chiping Li), NSERC Discovery and the National Research Council of Canada (Dr.\ Zekai Hong).

\bibliographystyle{proci}
\bibliography{references}

@article{lundstrom1969,
author = {Lundstrom, E. A.  and Oppenheim, A. K. },
title = {On the influence of non-steadiness on the thickness of the detonation wave},
journal = {Proceedings of the Royal Society of London. A. Mathematical and Physical Sciences},
volume = {310},
number = {1503},
pages = {463-478},
year = {1969},
doi = {10.1098/rspa.1969.0087}
}

@article{cheevers2022ignition,
author = {Cheevers, K. and Radulescu, M.I.},
title = {Ignition Behind Decaying Shock Waves: Detonation Cells},
journal = {preprint,arXiv:2211.05216v1 [physics.flu-dyn]},
doi = {10.48550/arXiv.2211.05216},
year = {2022}
}

@article{frederick2024,
title = {Reactive processes following transverse wave interaction},
journal = {Proceedings of the Combustion Institute},
volume = {40},
number = {1},
pages = {105552},
year = {2024},
issn = {1540-7489},
doi = {https://doi.org/10.1016/j.proci.2024.105552},
author = {Mark D. Frederick and Rohan M. Gejji and Joseph E. Shepherd and Carson D. Slabaugh}
}

@article{xiao2020role,
author = {Qiang Xiao and Matei I. Radulescu},
title = {Role of instability on the limits of laterally strained detonation waves},
journal = {Combustion and Flame},
volume = {220},
pages = {410-428},
year = {2020},
issn = {0010-2180},
doi = {10.1016/j.combustflame.2020.06.040}
}

@article{radulescu2018dynamic, 
title={Dynamics of detonations with a constant mean flow divergence},
volume={845},
DOI={10.1017/jfm.2018.244},
journal={Journal of Fluid Mechanics},
author={Radulescu, Matei I. and Borzou, Bijan}, 
year={2018},
pages={346-377}
}

@Article{radulescu2025gasdynamic,
 title={A gas-dynamic model for the dynamics of galloping detonations}, volume={1020}, DOI={10.1017/jfm.2025.10547}, journal={Journal of Fluid Mechanics}, author={Radulescu, Matei Ioan}, year={2025}, pages={R2}}

@article{frederick2023,
title = {Statistical analysis of detonation wave structure},
journal = {Proceedings of the Combustion Institute},
volume = {39},
number = {3},
pages = {2847-2854},
year = {2023},
issn = {1540-7489},
doi = {https://doi.org/10.1016/j.proci.2022.08.054},
url = {https://www.sciencedirect.com/science/article/pii/S1540748922003327},
author = {Mark D. Frederick and Rohan M. Gejji and Joseph E. Shepherd and Carson D. Slabaugh}
}

@article{radulescu2007hydrodynamic,
   author = {Radulescu, M. I. and Sharpe, G. J. and Law, C. K. and Lee, J. H. S.},
   title = {The hydrodynamic structure of unstable cellular detonations},
   journal = {Journal of Fluid Mechanics},
   volume = {580},
   pages = {31--81},
   DOI = {10.1017/s0022112007005046},
   year = {2007}
}

@article{crane2021,
title = {Geometric modeling and analysis of detonation cellular stability},
journal = {Proceedings of the Combustion Institute},
volume = {38},
number = {3},
pages = {3585-3593},
year = {2021},
issn = {1540-7489},
doi = {https://doi.org/10.1016/j.proci.2020.06.278},
url = {https://www.sciencedirect.com/science/article/pii/S1540748920303709},
author = {Jackson Crane and Xian Shi and Jonathan T. Lipkowicz and Andreas M. Kempf and Hai Wang}
}

@article{Frederick_Gejji_Shepherd_Slabaugh_2025, title={Reactive transverse waves in a near-limit detonation}, volume={1017}, DOI={10.1017/jfm.2025.10426}, journal={Journal of Fluid Mechanics}, author={Frederick, Mark D. and Gejji, Rohan M. and Shepherd, Joseph E. and Slabaugh, Carson D.}, year={2025}, pages={A10}}

@InProceedings{radulescu2025meanP,
   author = {Matei Radulescu and Farzane Zangene},
   title = {The mean pressure evolution in detonation cells},
   booktitle = {Proceedings of the 30th International Colloquium on Dynamics of Explosions and Reactive Systems},
   year = {2025},
   month = {July 27- August 1},
   address = {Ottawa, Canada},
   note = {paper 259}
}

@Article{denisov1959,
author = {Denisov, Y. N. and Troshin, Y. K.},
journal = {Proceedings of the Academy of Science USSR, Phys. Chem. Sec.},
year = {1959},
volume = {125},
pages = {217}
}

@article{lee1984,
   author = "Lee, J H S",
   title = "Dynamic Parameters of Gaseous Detonations", 
   journal= "Annual Review of Fluid Mechanics",
   year = "1984",
   volume = "16",
   number = "Volume 16, 1984",
   pages = "311-336",
   doi = "https://doi.org/10.1146/annurev.fl.16.010184.001523",
   url = "https://www.annualreviews.org/content/journals/10.1146/annurev.fl.16.010184.001523",
   publisher = "Annual Reviews",
   issn = "1545-4479",
   type = "Journal Article",
  }

@article{nakayama2013,
title = {Front shock behavior of stable curved detonation waves in rectangular-cross-section curved channels},
journal = {Proceedings of the Combustion Institute},
volume = {34},
number = {2},
pages = {1939-1947},
year = {2013},
issn = {1540-7489},
doi = {https://doi.org/10.1016/j.proci.2012.06.012},
url = {https://www.sciencedirect.com/science/article/pii/S1540748912001204},
author = {Hisahiro Nakayama and Jiro Kasahara and Akiko Matsuo and Ikkoh Funaki}
}

@article{vasiliev,
title = {Closed theoretical model of a detonation cell},
journal = {Acta Astronautica},
volume = {5},
number = {11},
pages = {983-996},
year = {1978},
issn = {0094-5765},
doi = {https://doi.org/10.1016/0094-5765(78)90004-8},
url = {https://www.sciencedirect.com/science/article/pii/0094576578900048},
author = {A.A. Vasiliev and Yu. Nikolaev}
}

@MastersThesis{lalchandani,
author = {Lalchandani, Sarthak},
title = {Modelling of Quasi Steady Detonations with Inert Confinement},
school = {Univesity of Ottawa},
year = {2022},
DOI = {10.20381/ruor-28571}
}

@article{takai,
title = {Study of detonation wave structure},
journal = {Symposium (International) on Combustion},
volume = {15},
number = {1},
pages = {69-78},
year = {1975},
issn = {0082-0784},
doi = {https://doi.org/10.1016/S0082-0784(75)80285-2},
url = {https://www.sciencedirect.com/science/article/pii/S0082078475802852},
author = {R. Takai and K. Yoneda and T. Hikita}
}

@InBook{bdzil2012,
author = {J. Bdzil and D.S. Stewart},
editor = {F. Zhang},
title = {Detonation Dynamics},
chapter = {Theory of Detonation Shock Dynamics},
publisher = {Springer},
year = {2012},
volume = {6},
series = {Shock Wave Science and Technology Reference Library}
}

@article{radulescu2002failure,
   author = {Radulescu, M. I. and Lee, J. H. S.},
   title = {The failure mechanism of gaseous detonations: Experiments in porous wall tubes},
   journal = {Combustion and Flame},
   volume = {131},
   number = {1-2},
   pages = {29--46},
   DOI = {10.1016/s0010-2180(02)00390-5},
   year = {2002}
}

@Article{hu2002analytical,
author = {Hu, X. and Zhang, D. and Jiang, Z.},
title = {Analytical study of idealized two-dimensional cellular detonations},
journal = {Shock Waves},
year = {2002},
pages = {475–480},
doi = {10.1007/s001930200124}
}

@article{MONNIER2023112996,
title = {From graph theory and geometric probabilities to a representative width for three-dimensional detonation cells},
journal = {Combustion and Flame},
volume = {256},
pages = {112996},
year = {2023},
issn = {0010-2180},
doi = {https://doi.org/10.1016/j.combustflame.2023.112996},
url = {https://www.sciencedirect.com/science/article/pii/S0010218023003723},
author = {Vianney Monnier and Pierre Vidal and Vincent Rodriguez and Ratiba Zitoun}
}

@article{JACKSON20132260,
title = {The influence of the cellular instability on lead shock evolution in weakly unstable detonation},
journal = {Combustion and Flame},
volume = {160},
number = {10},
pages = {2260-2274},
year = {2013},
issn = {0010-2180},
doi = {https://doi.org/10.1016/j.combustflame.2013.04.028},
url = {https://www.sciencedirect.com/science/article/pii/S0010218013001727},
author = {Scott I. Jackson and Mark Short}
}

@Article{watanabe2026,
author = {H. Watanabe and V. Robin and A. Chinnayya},
title = {Role of the deflagrative burning on the quenching limits for gaseous hydrogen detonation confined by an inert compressible layer},
journal = {Shock Waves},
year = {2026},
volume = {36},
number = {14},
doi = {10.1007/s00193-026-01262-8}
}

\appendix
\section{Derivation of the pdf of speed and residence time}
\subsection{Cell geometry and shock surface parametrization by $x$}
We now illustrate the procedure of obtaining the \textit{pdf} of the shock speed for a model 2D cell illustrated in Fig.\ \ref{fig:cellconstruct}. We assume the cell has length $L$ and width $\lambda$.  We assume the shock is flat, such that the shock position $x$ can parametrize the shock speed distribution in the cell.  We also assume that the cell has straight edges, such that the shock frontal area $A$ varies linearly with $x$. The evolution of the frontal area is given by 
\begin{align}
A(x)=\begin{cases}
      2 \lambda w \left( \frac{x}{L} \right) & \text{first half of cell}\\
      2 \lambda w \left( 1-\frac{x}{L} \right) & \text{second half of cell}
      \end{cases} 
\end{align}
where $w$ is the width of the 2D cell.  Given that the volume of the cell is
\begin{equation}
V_{cell}=\frac{\lambda L w}{2}
\end{equation}
we obtain
\begin{align}
\frac{A(x)}{V_{cell}} =\begin{cases}
      \frac{4}{L} \left( \frac{x}{L} \right) & \text{first half of cell}\\
      \frac{4}{L} \left( 1-\frac{x}{L} \right) & \text{second half of cell}
      \end{cases} \label{eq:areawithx}
\end{align}
\subsection{Shock kinematics and parametrization by $D$}
We assume that the shock decays exponentially in time
\begin{equation}
D=D_0\exp(-\alpha t) \label{eq:Dexp}
\end{equation}
where $\alpha$ is a modeling constant and $D_0$ is the shock speed at the start of the cell. Since $D=\mathrm{d} x_s/\mathrm{d}t$ with $x_s$ denoting the position of the shock, this expression can be integrated in order to show that the shock speed decays linearly with distance, yielding  
\begin{align}
D(x)=D_0 - \alpha x \label{eq:Dlinear}
\end{align}
Denoting the speed at the end of the cell as $D_1$, with $L$ being the cell length, we have
\begin{align}
\alpha = \frac{D_0-D_1}{L}
\end{align}
We can now write:
\begin{align}
\frac{x}{L} = \frac{D_0-D}{D_0-D_1}
\end{align}
which can be substituted in \eqref{eq:areawithx} to write
\begin{align}
\frac{A(x)}{V_{cell}} =\begin{cases}
      \frac{4}{L} \left( \frac{D_0 -D }{D_0-D_1} \right) & \text{first half of cell}\\
      \frac{4}{L} \left( \frac{D-D_1}{D_0 - D_1} \right) & \text{second half of cell}
      \end{cases} \label{eq:areawithd}
\end{align}
\subsection{PDF of shock speed}
We denote $\mathrm{pdf}(D) \delta D$ the volume of fluid swept by the shock while it is changing its speed from $D$ to $D + \delta D$ normalized by the total volume swept by the shock during an entire cell. The volume swept by the shock is due to the change in shock position from $x$ to $x+\delta x$ where $x$ is a coordinate normal to the shock.  If $A$ is the local surface of the shock interrogated, then we have
\begin{align}
 \mathrm{pdf}(D) |\delta D| = \frac{A }{V_{cell}}|\delta x|
\end{align}
The increments $\delta D$ and $\delta x$ are related by the first order Taylor expansion
\begin{align}
 \delta D = \frac{\mathrm{d} D }{\mathrm{d}x} \delta x
\end{align}
such that the desired probability density of shock speed is given by
\begin{align}
 \mathrm{pdf}(D) = \frac{A }{V_{cell}}|\frac{\mathrm{d} x }{\mathrm{d}D}| \label{eq:pdf}
\end{align}
For the assumed cell kinematics, the area variation is given by \eqref{eq:areawithd} while the derivative required for the change in variable is given differentiating \eqref{eq:Dlinear}, yielding
\begin{align}
|\frac{\mathrm{d} x }{\mathrm{d}D}| =  \frac{L}{D_0 - D_1}\label{eq:jac}
\end{align}
This results in the following triangular \textit{pdf} of shock speed 
\begin{align}
\mathrm{pdf}(D) =\begin{cases}
      4 \left( \frac{D_0 -D }{(D_0-D_1)^2} \right) & \text{first half of cell}\\
      4 \left( \frac{D-D_1}{(D_0 - D_1)^2} \right) & \text{second half of cell}
      \end{cases} \label{eq:pdf1}
\end{align}
It is more convenient to express the pdf in terms of the speed deficit at the end of the cell $\Delta D$ and the speed evaluated at the mid-cell location ($x = L/2$), i.e.,
\begin{align}
D_{mid} \equiv \frac{D_0+D_1}{2}\\
\Delta D \equiv D_{mid} - D_1
\end{align}
Using these new parameters, the distribution of shock speed can be written as 
\begin{align}
\mathrm{pdf}(D) =\begin{cases}
       \frac{D_{mid} +\Delta D - D }{(\Delta D)^2}  & \text{first half of cell}\\
       \frac{D-D_{mid} +\Delta D}{(\Delta D)^2}  & \text{second half of cell}
      \end{cases} \label{eq:pdf2}
\end{align}
Note that the mid-cell speed $D_{mid}$ is not exactly the mean propagation speed $\overline{D}$.  The mean propagation speed is given by 
\begin{align}
\overline{D} = \frac{L}{t_{cell}} 
\end{align} 
with the cell time $t_{cell}$ given by inverting \eqref{eq:Dexp}, i.e.,
\begin{equation}
t_{cell}=\frac{1}{\alpha}\ln\frac{D_0}{D_1} \label{eq:tcell}
\end{equation}
Let the parameter $\beta$ denote a non-dimensional speed deficit at the end of the cell:
\begin{equation}
\beta\equiv\frac{\Delta D}{D_{mid}}
\end{equation}
then \
\begin{equation}
\overline{D} = \frac{D_{mid} \beta}{\ln\frac{1+\beta}{1-\beta}}=D_{mid}(1+\mathcal{O}(\beta)) 
\end{equation}
We will assume that $\beta$ is small (typically 0.1 to 0.3) such that $\overline{D} \simeq D_{mid}$. 
With this approximation, instead of \eqref{eq:pdf2}, we can write the pdf of shock speed as parametrized by the mean propagation speed and $\beta$ in the following form
\begin{align}
\mathrm{pdf}(D) =\begin{cases}
       \left( 1+\beta- \frac{D}{\overline{D}} \right) \frac{1}{\beta^2 \overline{D}} & \text{first half of cell}\\
       \left( -1+\beta +\frac{D}{\overline{D}} \right)\frac{1}{\beta^2 \overline{D}}& \text{second half of cell}
      \end{cases} \label{eq:pdf3}
\end{align}
\subsection{pdf of non-reacted residence time in a cell}
Consider now a particle of gas shocked while propagating at speed $D$.  The time at which it is shocked inside the cell is obtained by inverting \eqref{eq:Dexp}, i.e., 
\begin{align}
t_s(D)=-\frac{1}{\alpha}\ln  \left(\frac{D}{D_0} \right)
\end{align}
Since all the gas is assumed to react close to the triple shock reflection at the end of the cell, the residence time, or \textit{age}, of a particle shocked at speed $D$ is thus
\begin{align}
t_{age}(D)=t_{cell}-t_s(D)
\end{align}
with the cell time $t_{cell}$ given by \eqref{eq:tcell}.  The residence time becomes
\begin{align}
t_{age}(D)=\frac{1}{\alpha} \ln  \left(\frac{D}{D_1} \right)
\end{align}
Using \eqref{eq:tcell}, we can re-write this expression as 
\begin{align}
t_{age}(D)= \frac{t_{cell}}{\ln  \left( \frac{D_0}{D_1}\right)}  \ln  \left(\frac{D}{D_1}\right) \label{eq:tresidence}
\end{align}

This expression can be inverted:
\begin{align}
\frac{D}{\overline{D}}=\frac{D_1}{\overline{D}}\exp \left(\ln  \left( \frac{D_0}{D_1} \right)\frac{t_{age}}{t_{cell}} \right)
\end{align}
For small $\beta$, it becomes
\begin{align}
\frac{D}{\overline{D}}\simeq\ (1-\beta)\exp \left( \frac{2 \beta}{t_{cell}}t_{age} \right) \label{eq:Dlife}
\end{align}
The pdf of the residence time can now be obtained from the pdf of shock speed:
\begin{align}
\mathrm{pdf}(t_{age})= \mathrm{pdf}\left(D(t_{age}) \right) |\frac{\mathrm{d} D}{\mathrm{d} t_{age}}|
\end{align}
with the derivative evaluated from \eqref{eq:tresidence} as 
\begin{align}
\frac{\mathrm{d} D}{\mathrm{d} t_{age}}&=\alpha D\\
&=\frac{D}{t_{cell}} \ln  \left( \frac{D_0}{D_1} \right)\\
&\simeq \frac{2 \beta D}{t_{cell}}
\end{align}
Using \eqref{eq:pdf3}, we obtain the desired pdf of residence time as non-reacted gas behind the shock:
\begin{align}
\mathrm{pdf}(t_{age}) =\begin{cases}
       \left( 1+\beta- \frac{D}{\overline{D}} \right) \frac{D}{\overline{D}} \frac{2}{\beta t_{cell}} & \text{first half of cell}\\
       \left( -1+\beta +\frac{D}{\overline{D}} \right)\frac{D}{\overline{D}} \frac{2}{\beta t_{cell}}& \text{second half of cell}
      \end{cases} \label{eq:pdf4}
\end{align}
together with $D/\overline{D}=f(\beta, t_{age}/t_{cell})$ given by \eqref{eq:Dlife}.
\section{Detonation speed predictions in linear-scale plots}
\begin{figure}
\centering
\includegraphics[width=0.7\textwidth]{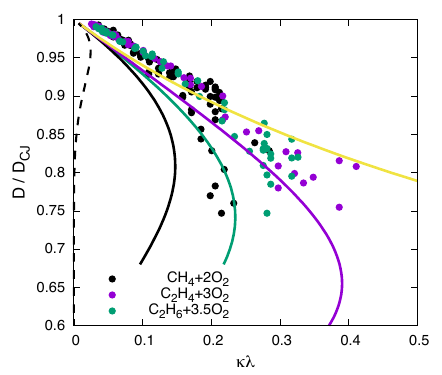}
\caption{Detonation speed dependence on mean curvature $\kappa \lambda$ obtained experimentally by \cite{xiao2020role}.  Yellow line is the model prediction using the closed form model of the present paper and fixed cellular time.  Colour coded lines are model predictions with state sensitive cellular times according to \eqref{eq:tcelldiff} for diffusively controlled pocket burn-out; broken line is the model prediction for CH$_4+2$O$_2$ according to \eqref{eq:tcellshock} for auto-ignition controlled pocket burn-out.}
\label{fig:qiang2linear}
\end{figure}

\begin{figure}
\centering
\includegraphics[width=0.7\textwidth]{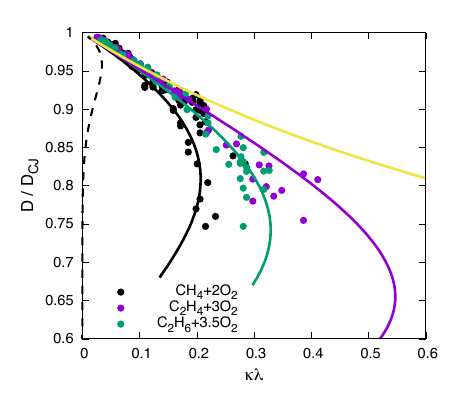}
\caption{Detonation speed dependence on mean curvature $\kappa \lambda$ obtained experimentally by \cite{xiao2020role}.  Model lines same as Fig.\ \ref{fig:qiang2linear} with the rate being increased by a factor of 1.4.}
\label{fig:correctedlinear}
\end{figure}

It is of interest to report Figures \ref{fig:qiang2} and \ref{fig:corrected} in linear coordinates to assess the model performance close to the propagation limits. The linear scale version of these two figures is shown in Figures \ref{fig:qiang2linear} and \ref{fig:correctedlinear}.

\newpage

\small
\baselineskip 10pt


\end{document}